\documentclass{emulateapj}
\usepackage{graphicx}
\usepackage{apjfonts}

\newcommand{\sauron}{{\tt SAURON}}
\newcommand{\SN}{\ensuremath{\mathrm{S/N}}}
\newcommand{\ud}{\mathrm{d}} % for integrals dx, dy, ...
\newcommand{\kms}{\hbox{km s$^{-1}$}}

\begin{document}

\title{Parametric Recovery of Line-of-Sight Velocity Distributions from\\ Absorption-Line Spectra of Galaxies via Penalized Likelihood}

\slugcomment{To appear in PASP, v.~116, 2004 February}
\shorttitle{Penalized Pixel Fitting}
\shortauthors{Cappellari \& Emsellem}

\author{Michele Cappellari}
\affil{Leiden Observatory, Postbus 9513, 2300 RA Leiden, The Netherlands}
\email{cappellari@strw.leidenuniv.nl}
\and
\author{Eric Emsellem}
\affil{Centre de Recherche Astronomique de Lyon, 9~Avenue Charles Andr\'e, 69230 Saint Genis Laval, France}
\email{emsellem@obs.univ-lyon1.fr}

\begin{abstract}
We investigate the accuracy of the parametric recovery of the line-of-sight velocity distribution (LOSVD) of the stars in a galaxy, while working in pixel space. Problems appear when the data have a low signal-to-noise ratio, or the observed LOSVD is not well sampled by the data. We propose a simple solution based on maximum penalized likelihood and we apply it to the common situation in which the LOSVD is described by a Gauss-Hermite series. We compare different techniques by extracting the stellar kinematics from observations of the barred lenticular galaxy NGC~3384 obtained with the \sauron\ integral-field spectrograph.
\end{abstract}


\section{Introduction}

The dynamics of stars in a galaxy is uniquely defined by the distribution function and the potential in which the stars move. From the many stable equilibrium configurations for collisionless systems which can in principle be constructed, only some of them are actually observed in our Universe. This constitutes a `fossil record' of galaxy formation. It is still not clear what observable constraints are needed to recover both the distribution function and the gravitational potential of a stellar system \citep[but see, e.g.,][for spherical geometry]{dej92}, but simple dimensionality arguments imply that this should not be possible without the knowledge of the full line-of-sight velocity distributions (LOSVD) at all spatial positions on the galaxy image on the sky. It is therefore useful to explore how the LOSVD can be recovered from the observations.

Considering galaxies as pure stellar systems, the spectrum observed at a certain sky position is a (luminosity-weighted) sum of individual stellar spectra redshifted according to their line-of-sight velocities. If one makes the assumption that the spectrum of all stars is given by a single template, then it simply reduces to the convolution between that spectrum and the LOSVD, which can then be retrieved by solving the inverse problem, i.e., deconvolving the spectra using the template. Deconvolution is an intrinsically ill-conditioned problem, which amplifies noise and measurement errors. Due to the special care required in the inversion many techniques have been developed in the last 30 years to recover the LOSVD from the data. The evolution in the methods has been mainly driven by the gradual improvement in the observational techniques and the signal-to-noise ratio (\SN) of the data, and by the steady increase in the available computational speed.

Early methods were mostly using Fourier based techniques, which allowed the LOSVD to be recovered quickly, from a deconvolution process, and in some cases included techniques to reduce the effect of template mismatch \citep{sim74,sar77,ton79,fra88,ben90,sta95}. More recently methods have shifted towards fitting the LOSVD directly in pixel space \citep{rix92,kui93,van94,sah94,mer97,geb00,kel00}. The reasons for this are that (i) in pixel space it becomes easy to exclude gas emission lines or bad pixels from the fit, and take continuum matching directly into account; (ii) current computers can accommodate the larger computational cost involved; (iii) the availability of libraries with high spectral resolution stellar and galaxy spectra allows the template to be carefully matched to the observed galaxy spectrum \citep[e.g.,][]{ems04}. See \citet{deb03} for a more detailed historical overview of the various methods.

The different techniques can be further subdivided according to whether the LOSVD is derived in a non-parametric way (in practice computed on a small set of discrete values) or parametrically, as a function of a limited number of parameters. In the latter case the Gauss-Hermite parametrization by \citet{van93} and \citet{ger93} is essentially always adopted \citep[but see, e.g.,][for an alternative]{ZP96}. However even when the LOSVD is determined non-parametrically, the Gauss-Hermite parameters are still generally used to present the result in an easily understandable way.

In this paper we study again the problem of recovering, while working in pixel space, a LOSVD described by the Gauss-Hermite parametrization. Compared to the non-parametric case, the process and the estimation of measurement errors are simplified.
In Sec.~\ref{formulation} we describe the general problem. In Sec.~\ref{discussion} we discuss different approaches to the extraction, we find that special care has to be taken when the LOSVD is undersampled by the data or the \SN\ is low, and we present a solution based on the maximum penalized likelihood formalism. In Sec.~\ref{conclusion} we draw some conclusions.

\section{Formulation of the problem}
\label{formulation}

The parametric recovery of the LOSVD in pixel space starts with creating a model galaxy spectrum $G_{\rm mod}(x)$, by convolving a template spectrum $T(x)$ by a parametrized LOSVD. Both the object and the template spectra are rebinned in wavelength to a linear scale in $x=\ln\lambda$, while usually preserving the number of spectral pixels. The best-fitting parameters of the LOSVD are determined by minimizing the $\chi^2$, which measures the agreement between the model and the observed galaxy spectrum $G(x)$, over the set of $N$ good pixels:
\begin{equation}
    \chi^2 = \sum_{n=1}^N r_n^2
    \label{chi2}
\end{equation}
where the residuals are defined as
\begin{equation}
    r_n = \frac{G_{\rm mod}(x_n)-G(x_n)}{\Delta G(x_n)},
    \label{resid}
\end{equation}
with $\Delta G(x_n)$ the measurement error on $G(x_n)$.

More specifically the following model is adopted for the galaxy spectrum:
\begin{equation}
G_{\rm mod}(x) =
    \sum_{k=1}^{K} w_k [B \ast T_k](x) + \sum_{l=0}^{L} b_l \mathcal{P}_l(x) \qquad w_k\ge0,
    \label{model}
\end{equation}
where $T_k$ is in general a library of $K$ galaxy or stellar templates, $B(x)=\mathcal{L}(c x)$ is the broadening function, with $\mathcal{L}(v)$ the LOSVD, $c$ is the speed of light and $\ast$ denotes convolution.
The $\mathcal{P}_l(x)$ are here chosen to be the Legendre polynomials of order $l$ and account for low frequency differences in shape between the galaxy and the templates. For each given $\mathcal{L}(v)$, the optimization of $\chi^2$ is a bounded-variables linear least-squares problem for the weights $(w_1,\ldots,w_K,b_0,\ldots,b_L)$ which can be solved, e.g., with the specific BVLS algorithm by \citet{law95}, or as a quadratic programming problem. Here we are interested in the determination of the parameters defining $\mathcal{L}(v)$ and in what follows we will assume that the weights of Eq.~(\ref{model}) are always optimized in this way. Multiplicative polynomials can also be included in the fit \citep[see][]{kel00}, without affecting the discussion that follows.

Following \citet{van93} and \citet{ger93} it has become standard to expand the LOSVD as a Gauss-Hermite series
\begin{equation}
\mathcal{L}(v)=\frac{e^{-(1/2)y^2}}{\sigma\sqrt{2\pi}}
    \left[ 1 + \sum_{m=3}^M h_m H_m(y) \right]
    \label{losvd}
\end{equation}
where $y=(v-V)/\sigma$ and the $H_m$ are the Hermite polynomials.
With these definitions the minimization of the $\chi^2$ in Eq.~(\ref{chi2}) is a nonlinear least-squares optimization problem for the $M$ parameters $(V,\sigma,h_3,\ldots,h_M)$. Least-squares problems can be solved much more efficiently than general ones, by using specific algorithms which require the user to provide the  residuals $r_n$ of Eq.~(\ref{resid}), to compute explicitly the Hessian matrix of the $\chi^2$ merit function \citep[see, e.g.,][\S 15.5]{pre92}. Here we will use the MINPACK\footnote{We used an IDL porting of the code by Craig B.\ Markwardt, available from the Web page \url{http://cow.physics.wisc.edu/$\sim$craigm/idl/}} implementation \citep{mor80} of the Levenberg-Marquardt method for nonlinear least-squares problems.

\section{Discussion}
\label{discussion}

In this section we compare three different approaches to the determination of the best-fitting parameters of the LOSVD in Eq.~(\ref{losvd}). We explain the limitations of the different methods and finally select the last one as the optimal choice. We do not address here the template mismatch issue, which we assume is minimized by the choice of an optimal library of templates in Eq.~(\ref{model}) as in \citet{ems04}.

\begin{figure}
\epsscale{1.15}
\plotone{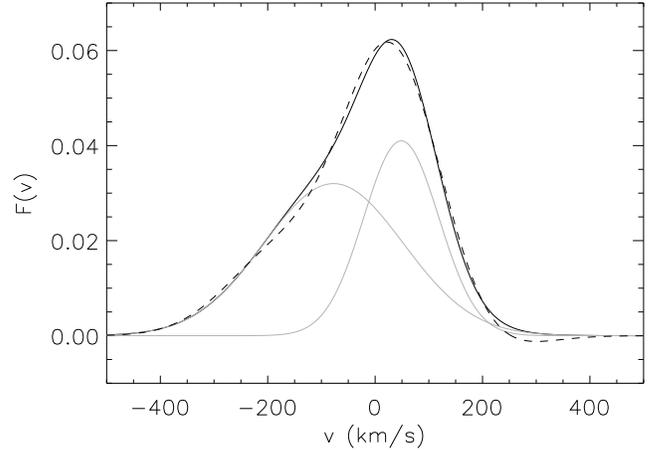}
\caption{Double Gaussian representation (Eq.~\ref{double_gauss}) of the LOSVD observed by \citet{sco95} at 17\arcsec\ along the major axis of the S0 galaxy NGC~3115 (solid line). The two individual Gaussian components are shown with the gray lines, while the best fitting fourth order Gauss-Hermite parametrization is plotted with the dashed line.\label{sb_losvd}}
\end{figure}

To explain the characteristics of the three methods, we will use each of them to extract a realistic but known LOSVD, observed by \citet{sco95} at $r=17\arcsec$ along the major axis of the S0 galaxy NGC~3115 (their Fig.~A.2). This LOSVD (Fig.~\ref{sb_losvd}) is representative of the one observed in a number of galaxies, and can be very well described by a double Gaussian parametrization
\begin{equation}
    F(v) = \sum_{j=1}^{2} I_j\, \exp\left[\frac{(v-V_j)^2}{2\;\sigma_j^2}\right],
    \label{double_gauss}
\end{equation}
with parameters $I_1=0.041$, $I_2=0.032$, $V_1=48.7$ \kms, $V_2=-77.3$ \kms, $\sigma_1=70.0$ \kms\ and $\sigma_2=130.0$ \kms\ (the LOSVD was shifted to zero mean velocity).

The best fit to $F(v)$ using a fourth order Gauss-Hermite series (Eq.~\ref{losvd}) is obtained with parameters $V=0.0$ \kms, $\sigma=114.8$ \kms, $h_3=-0.150$ and $h_4=0.036$, while the best fitting Gaussian has $V=-2.3$ \kms\ and $\sigma=118.9$ \kms.

\subsection{Fitting $(V,\sigma)$ first}

In an ideal situation where the LOSVD is perfectly sampled by the data, the $(h_3,\ldots,h_M)$ parameters of the LOSVD (Eq.~\ref{losvd}) are essentially uncorrelated to $(V,\sigma)$. Therefore one expects the best-fitting parameters to change little, irrespective of whether $(V,\sigma)$ are fitted first or together with the other parameters. To lowest order the difference between the two approaches is given by Eq.~(10) of \citet{van93}.

An advantage of fitting $(V,\sigma)$ first is that the Gauss-Hermite parameters that one obtains, coincide with the true Gauss-Hermite moments integrated over the LOSVD \citep[see][for a detailed discussion]{van93}. This also means that the value of the parameters $(h_3,\ldots,h_M)$ do not depend on the adopted number $M$ of terms in the expansion. This precise relation between the LOSVD and the extracted kinematical parameters can be very useful for the dynamical modeling \citep[e.g.][]{rix97}.

Moreover when only a single template is adopted ($K=1$ in Eq.~\ref{model}), for each pair of $(V,\sigma)$ values the best fitting value of all the remaining parameters $(w_1,h_3,\ldots,h_M,b_0,\ldots,b_L)$ can be determined linearly.
These ideas led \citet{van94} to design an efficient method to fit the parameters of the LOSVD in pixel space, where $(V,\sigma)$ are fitted first and the remaining parameters are linearly expanded at the best fitting $(V,\sigma)$ solution.

For extracting the stellar kinematics of the 72 galaxies of the \sauron\ survey \citep{dez02} we needed to operate in pixel space to be able to deal with the contamination due to emission-line gas present in most of the observed objects \citep{ems04}. The speed of \citet{van94} pixel fitting algorithm was an attractive feature, given the large number ($\sim200.000$) of independent spectra from the survey and the need to compute accurate errors by Monte Carlo simulations.
However the observed LOSVDs are not always well sampled by the \sauron\ pixel scale of $\ud v\approx60$ \kms. The Nyquist critical frequency for $h_4$ is $\sim\sigma/2$. This means that undersampling becomes a problem for the derivation of the first four Gauss-Hermite moments of the LOSVD when the observed velocity dispersion is of the order of $\sim2$ pixels (for \sauron\ $120$ \kms).

\begin{figure}
\epsscale{1.15}
\plotone{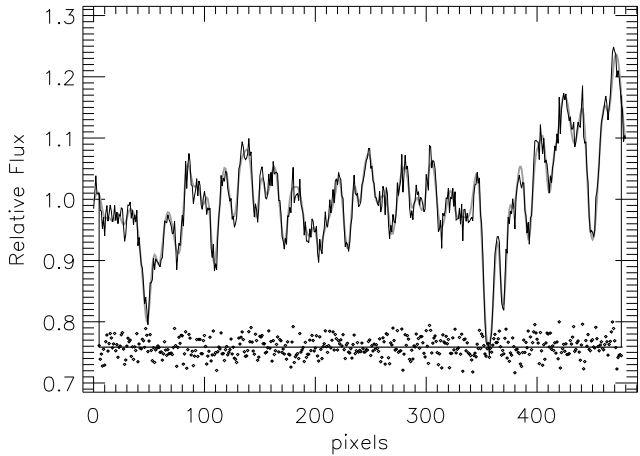}
\caption{Logarithmically rebinned galaxy model spectrum (thin noisy line) convolved with the LOSVD of Fig.~\ref{sb_losvd} and with added Poissonian noise at a $\SN=60$. The dots represent the residual difference between this model and the best fitting template (thick gray line). The spectrum covers the wavelength range 4800-5380 \AA, with a pixel scale of 60 \kms, and includes prominent absorption features of $H\beta$, Mg$\,b$ and Fe5270.\label{spectrum}}
\end{figure}

To test the accuracy of the recovery of the Gauss-Hermite moments using this method, when undersampling becomes significant, we first created a realistic model spectrum by fitting a (logarithmically rebinned) library of \citet{vaz99} galaxy model spectra to the average \sauron\ spectrum of the barred lenticular galaxy NGC~3384. This spectrum was oversampled by smoothly interpolating it on a 30 times finer spectral grid and was subsequently convolved with the LOSVD of Fig.~\ref{sb_losvd}. It was integrated over the \sauron\ pixels, in the wavelength range 4800-5380 \AA, and Poissonian noise was finally added to represent the simulated galaxy spectrum $G(x)$. The template $T(x)$ was obtained by integrating the original oversampled spectrum over the \sauron\ pixels. In Fig.~\ref{spectrum} we show an example of such a spectrum, with noise added at $\SN=60$.

We subsequently tried to recover the true Gauss-Hermite moments, by fitting $(V,\sigma)$ first and expanding $(h_3,\ldots,h_M)$ linearly at the best fitting $(V,\sigma)$ location. To check our results we computed the true moments $(\hat{h}_3,\hat{h}_4)$, by analytically integrating over the LOSVD \citep[Eq.~8 of][]{van93}, using $(V,\sigma)$ of the best fitting Gaussian computed in Sec.~\ref{discussion}. We found $\hat{h}_3=-0.144$ and $\hat{h}_4=0.029$, which as expected are very close to the best fitting parameters computed before.

\begin{figure}
\epsscale{1.15}
\plotone{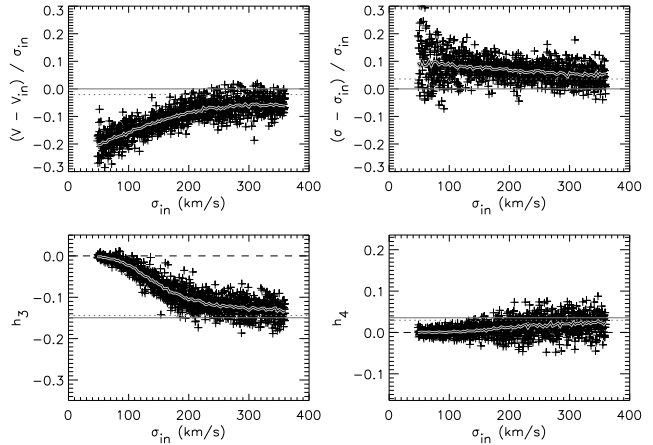}
\caption{Fitting $(V,\sigma)$ first. Recovery of the stellar kinematics from a model galaxy spectrum convolved with the LOSVD of Fig.~\ref{sb_losvd}. The shape of the LOSVD was held fixed, while the velocity scale was varied so that the $\sigma_{\rm in}$ of the best fitting Gauss-Hermite series varied in the range 48--360 \kms, corresponding to 0.8--6 pixels. In these measurements $V$ and $\sigma$ were fitted first (nonlinearly) and only then $h_3$ and $h_4$ were fitted (linearly) with $V$ and $\sigma$ fixed to the optimal values. In each panel the thick dotted line represents the true Gauss-Hermite moments, that the method is trying to recover, the thick solid line indicates the value of the best fitting Gauss-Hermite parameters, while the crosses represent the actually measured values, as a function of the input $\sigma_{\rm in}$, for 1000 different Monte Carlo realizations with $\SN=60$. The thick noisy line shows the measured values for 100 realizations with a very high $\SN=600$. The recovered moments start becoming biased when $\sigma_{\rm in}\la240$ (4 pixels), and this method becomes essentially insensitive to any deviation from a Gaussian when $\sigma_{\rm in}\la120$ \kms\ (2 pixels), at any \SN.\label{v_sigma}}
\end{figure}

The result of this test, both for a $\SN=60$ and $\SN=600$, is shown in Fig.~\ref{v_sigma}. Here the shape of the input LOSVD was held fixed, but the velocity scale was varied so that the $\sigma_{\rm in}$ of the best fitting Gauss-Hermite series varied in the range 48--360 \kms, corresponding to 0.8--6 \sauron\ pixels, for 1000 different Monte Carlo realizations. By construction, for $\sigma_{\rm in}=114.8$ \kms\ the LOSVD reduces to the one presented in Fig.~\ref{sb_losvd}. The recovery of the true moments is {\em strongly} biased when $\sigma_{\rm in}\la240$ \kms\ (4 pixels). This can be understood by the fact that, when the LOSVD is not very well sampled, an asymmetry in the profile can be compensated, during the Gaussian fit, by a small $V$ shift, while a symmetric deviation from a Gaussian is nearly equivalent to a change of $\sigma$. The bias in the recovery of the input parameters remains unchanged even for a noise-free model spectrum, although of course the scatter in the values disappears. And no improvement is observed if the template is oversampled before convolving it with a well sampled LOSVD in Eq.~(\ref{model}). These results are representative of what we found with different realistic input LOSVDs.

What was seen in the simulation is also found on real \sauron\ data. The approach of fitting $(V,\sigma)$ first, becomes insensitive to any deviation from a Gaussian LOSVD, when the observed velocity dispersion is low ($\la120$ \kms, or 2 pixels). As a practical example this method gives the misleading impression that the LOSVD in the nucleus of a low-dispersion object like NGC~3384, containing a fast-rotating disk-like structure, is consistent with being essentially symmetric, as, e.g., the central LOSVD of the high-dispersion non-rotating giant elliptical galaxy M~87. This apparent similarity of the overall shape of LOSVD in the two different galaxies is an artefact of the extraction method, and we will show that it is still possible to recover, with a different method, the strong asymmetry of the LOSVD from the \sauron\ data of NGC~3384, even at low dispersion.

\subsection{Fitting all parameters simultaneously}
\label{together}

The bias present in the recovery of the Gauss-Hermite moments of an undersampled LOSVD, in the previous section, was mainly due to the fact that $(V,\sigma)$ were not fitted simultaneously with the other LOSVD parameters. The fit should then become generally unbiased when all $(V,\sigma,h_3,\ldots,h_M)$ parameters are varied simultaneously to optimize the fit. In this way one should recover the best fitting parameters of the Gauss-Hermite series, which do not precisely coincide with the true Gauss-Hermite moments of the LOSVD. By definition this method will provide the lowest $\chi^2$ in the fit, and for this reason it was also originally chosen by \citet{van93}.

\begin{figure}
\epsscale{1.15}
\plotone{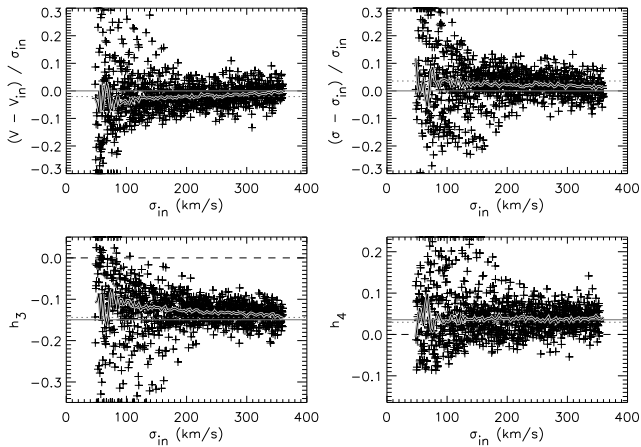}
\caption{Same as in Fig.~\ref{v_sigma}, but in this plot the four parameters $(V,\sigma,h_3,h_4)$ were all fitted simultaneously (nonlinearly). Now the measurements are supposed to reproduce the best-fitting Gauss-Hermite parameters, which are indicated by the thick solid line. The measurements are generally unbiased, but no reliable $h_3$ and $h_4$ measurements can be obtained at $\SN=60$ for $\sigma_{\rm in}\la120$ \kms\ (2 pixels). At low $\sigma_{\rm in}$ the scatter in $V$ and $\sigma$ increases due to their correlation with $h_3$ and $h_4$ respectively. Moreover the measured $\sigma$ tends to be systematically lower than the true value, while $h_4$ is correspondingly too high. In the case of very high \SN\ the strong asymmetry of the LOSVD can be recovered, with a modest bias, down to the smallest $\sigma_{\rm in}$ values.\label{fit_all}}
\end{figure}

In Fig.~\ref{fit_all} we repeated the experiment of the previous section while fitting all the above nonlinear parameters together. As expected the solution is now almost unbiased and the measurements tend to be spread around the true input values, for a wider $\sigma_{\rm in}$ range. However, when $\sigma_{\rm in}\la120$ \kms\ (2 pixels) and the LOSVD is correspondingly not well sampled, the spectrum does not contain enough information to constrain all the free parameters and the scatter in the recovered values increases dramatically.

\begin{figure}
\epsscale{1.15}
\plotone{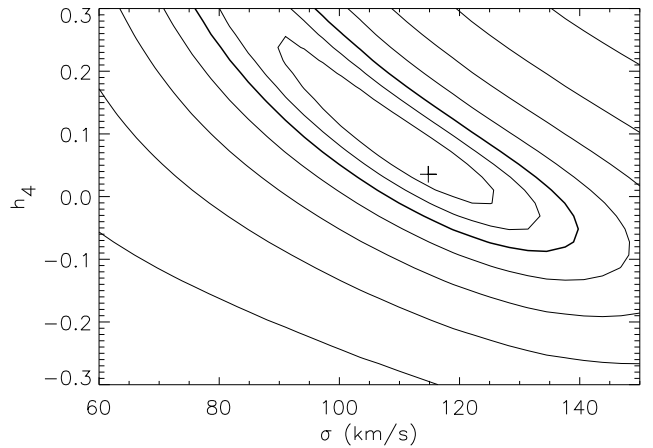}
\caption{Contours of the $\Delta\chi^2=\chi^2-\chi^2_{\rm min}$ which measure the agreement between a simulated noiseless input galaxy spectrum $G(x)$ and its best fitting model $G_{\rm mod}(x)$. As in Fig.~\ref{fit_all} $G(x)$ was convolved with the LOSVD of Fig.~\ref{sb_losvd}. The $\Delta\chi^2$ is plotted as a function of $\sigma$ and $h_4$ of the LOSVD of $G_{\rm mod}(x)$, while $V$ and $h_3$ are kept fixed to the known best fitting values (Sec.~\ref{discussion}). The plus sign indicates the location of the known best fitting $(\sigma,h_4)$. The three lowest $\Delta\chi^2$ levels correspond to confidence levels of 1, 2 and $3\sigma$ (thick line) respectively, for a $\SN=60$, while other levels are separated by factors of 2. Note the narrow valley of nearly constant $\chi^2$, for decreasing $\sigma$ and increasing $h_4$, which is due to the undersampling of the LOSVD. At this $\sigma_{\rm in}$ a very high \SN\ is needed for the fit to converge to the true minimum.
\label{chi2plot}}
\end{figure}

Moreover, at low $\sigma_{\rm in}$, the $\sigma$ and $h_4$ values are not unbiased any more, not being symmetrically distributed around the known input values. The reason for this asymmetry can be understood by looking at the shape of the $\chi^2$ contours, which measure the agreement between the input spectrum $G(x)$ and its best fitting model $G_{\rm mod}(x)$. In Fig.~\ref{chi2plot} we plot the $\Delta\chi^2=\chi^2-\chi^2_{\rm min}$ contours for fits obtained by convolving  $G(x)$ with the LOSVD of Fig.~\ref{sb_losvd}, while keeping $V$ and $h_3$ fixed to the known best fitting values (Sec.~\ref{discussion}) and varying $\sigma$ and $h_4$ of $G_{\rm mod}(x)$. The narrow curved valley of nearly constant $\chi^2$, for decreasing $\sigma$ and increasing $h_4$ from the location of the minimum $\chi^2_{\rm min}$, is due to the strong correlation between these two parameters, which is caused by the undersampling, and produces the effects seen in Fig.~\ref{fit_all}.

When the errors in the measured parameters are large, one has to make a scatter versus bias tradeoff decision. Usually when the data are unable to tightly constrain a large number of parameters one prefers to retain in the fit only the parameters that are required to significantly decrease the $\chi^2$. For this reason, while fitting a parametric LOSVD, it is common practice to reduce the fit to $(V,\sigma)$ at low \SN: although the fit may be biased by the lack of freedom in the model, an additional flexibility would not lead to meaningful measurements, and structures and trends in the $(V,\sigma)$ values can still be detected due to the decreased scatter.

Another problem of the large scatter that is present when fitting all parameters together is specific to the two-dimensional (2D) kinematical measurements obtained with an integral-field spectrograph. When presenting kinematics maps there is no easy way to also attach errors to the displayed values, and it becomes difficult to estimate from the maps when some structure is real and when it is due to noise. For this reason it would be preferable to display on the kinematics maps only the features that are statistically significant.

\subsection{Penalized pixel fitting}

The LOSVD of galaxies is generally well reproduced by a Gaussian \citep[e.g.,][]{ben94}. For this reason one would like to use a technique to fit the LOSVD in which the solution is free to reproduce the details of the actual profile when the \SN\ is high, but where the solution tends to a Gaussian shape in case the \SN\ is low. This was done for the case the LOSVD is described by a non-parametric function, by using the maximum penalized likelihood formalism \citep[e.g.,][]{mer97}, and we refer to that paper for details. Here we apply the formalism to the case where the LOSVD is parametrically expanded as a Gauss-Hermite series. We will show that this leads to a much simpler and faster implementation than for the general case.

The idea is to fit the parameters $(V,\sigma,h_3,\ldots,h_M)$ simultaneously as in Sec.~\ref{together}, but to add an adjustable penalty term to the $\chi^2$, to bias the solution towards a Gaussian shape, when the higher moments are unconstrained by the data, so that the penalized $\chi^2$ becomes:
\begin{equation}
    \chi_{\rm p}^2=\chi^2 + \alpha \mathcal{P}.
    \label{penchi2}
\end{equation}
A natural form for the penalty function  $\mathcal{P}$ is given by the integrated square deviation of the line profile $\mathcal{L}(v)$ from its best fitting Gaussian $\mathcal{G}(v)$:
\begin{equation}
\mathcal{D}^2=\frac{\int_{-\infty}^{\infty} [\mathcal{L}(v)-\mathcal{G}(v)]^2 \ud v}
{\int_{-\infty}^{\infty} \mathcal{G}(v)^2 \ud v}.
\label{rms}
\end{equation}
This penalty does not suppress noisy solutions, which are already excluded by the use of a low order parametric expansion for  $\mathcal{L}(v)$.
It was shown by \citet{van93} that in the case where $\mathcal{L}(v)$  has the form of Eq.~(\ref{losvd}) then Eq.~(\ref{rms}) is well approximated by:
\begin{equation}
\mathcal{D}^2\approx\sum_{m=3}^{M} h_m^2.
\label{d2}
\end{equation}

In principle one could then define the penalty as $\mathcal{P}=\mathcal{D}^2$ and optimize $\chi_{\rm p}^2$ (Eq.~\ref{penchi2}). In practice however this is not desirable for two reasons:
\begin{itemize}
\item as explained in Sec.~\ref{formulation}, it is computationally much more efficient to minimize the residuals $r_n$ (Eq.~\ref{resid}) instead of explicitly compute the $\chi^2$;
\item one needs a way to automatically adjust the penalty factor $\alpha$ according to the $\chi^2$ of the observed fit.
\end{itemize}

We found that a simple and effective solution to these problems consists of using the following perturbed residuals as input to the nonlinear least-squares optimizer:
\begin{equation}
r'_n = r_n + \lambda\; \sigma(\mathbf{r})\; \mathcal{D},
\label{pert}
\end{equation}
where the variance is defined as
\begin{equation}
\sigma^2(\mathbf{r})=\frac{1}{N}\sum_{n=1}^N r_n^2.
\label{var}
\end{equation}
The qualitative interpretation of this formula is that a deviation $\mathcal{D}$ of the LOSVD from a Gaussian shape will be accepted as an improvement of the fit, only if it is able to correspondingly decrease the scatter $\sigma(\mathbf{r})$ by an amount related to $\mathcal{D}$.
To quantify one can compute the objective function of the fit
\begin{equation}
\chi_{\rm p}^2=\sum_{n=1}^N r_n^2\; +\; 2\lambda \sigma(\mathbf{r}) \mathcal{D} \sum_{n=1}^N r_n\; +\; N\; [\lambda \sigma(\mathbf{r}) \mathcal{D}]^2.
\end{equation}
The sum of the residuals in the second term is zero by construction, due to the fact that the weights are optimized for given $\mathcal{L}(v)$ (Sec.~\ref{formulation}; this is required for this form of perturbation to work). Considering the definition of variance (Eq.~\ref{var}) one can finally write
\begin{equation}
\chi_{\rm p}^2\;=\;\chi^2\; (1+\lambda^2 \mathcal{D}^2),
\label{chi2p}
\end{equation}
which is of the desired form (\ref{penchi2}) with $\alpha=\lambda^2 \chi^2$ automatically scaled according to the $\chi^2$ of the fit. In practice $\sigma(\mathbf{r})$ in Eq.~(\ref{pert}) is computed using a robust biweight estimator \citep{hoa83}, so that Eq.~(\ref{chi2p}) is only valid as an approximation. One can see from this formula that, with $\lambda=1$, a deviation  $\mathcal{D}=10\%$ of the LOSVD from a Gaussian (e.g., $h_3=0.1$ and $h_m=0$) requires a corresponding decrease in the scatter $\sigma(\mathbf{r})$ of the unperturbed residuals by  more than $(1+\mathcal{D}^2)^{1/2}=0.5\%$ to be accepted by the optimization routine as an improvement of the fit.

A statistical interpretation of Eq.~(\ref{chi2p}) can be obtained by considering that for a good fit $\chi^2\sim N$. For a given $\lambda$ a deviation $\mathcal{D}$ from a Gaussian needs to decrease $\chi^2$ by more than $\Delta\chi^2\sim N\lambda^2\mathcal{D}^2$ to be accepted, and this variation can be associated to a specific confidence level \citep[see, e.g.,][\S 15.6]{pre92} at which a Gaussian shape is excluded. Although this formula may serve as a guideline, we believe it is safer to test a choice of $\lambda$ using simulations as the ones presented.

\begin{figure}
\epsscale{1.15}
\plotone{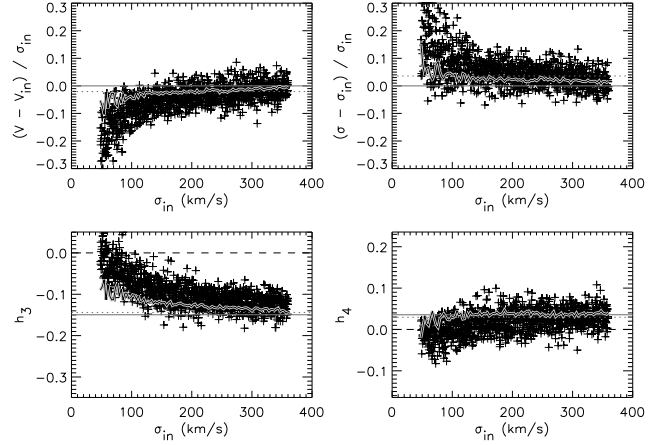}
\caption{Penalized pixel fitting. Same as in Fig.~\ref{fit_all} but using the new penalized parametric pixel fitting method with $\lambda=0.7$. At $\SN=60$ the bias on the parameters is now significant only when $\sigma_{\rm in}$ goes below about 120 \kms\ (2 pixels). But more importantly the fit tends to converge to the true solution even for low $\sigma_{\rm in}$ at the higher $\SN=600$. \label{penalized}}
\end{figure}

A test of the application of Eq.~(\ref{pert}) to the same model spectra of Sec.~\ref{together} is shown in Fig.~\ref{penalized}. For the same $\SN=60$ as before, using $\lambda=0.7$, the measurements are now essentially unbiased when $\sigma_{\rm in}\ga120$ \kms\ (2 pixels). For lower input dispersion the LOSVD converges towards a Gaussian and, as expected, the bias becomes similar to Fig.~\ref{v_sigma}. However the {\em crucial} difference with the previous case consists of the fact that for higher $\SN=600$ the solution tends to converge to the known best fitting values, due to the fact that the penalty scales with the $\chi^2$.

\begin{figure*}
\epsscale{1.03}
\plotone{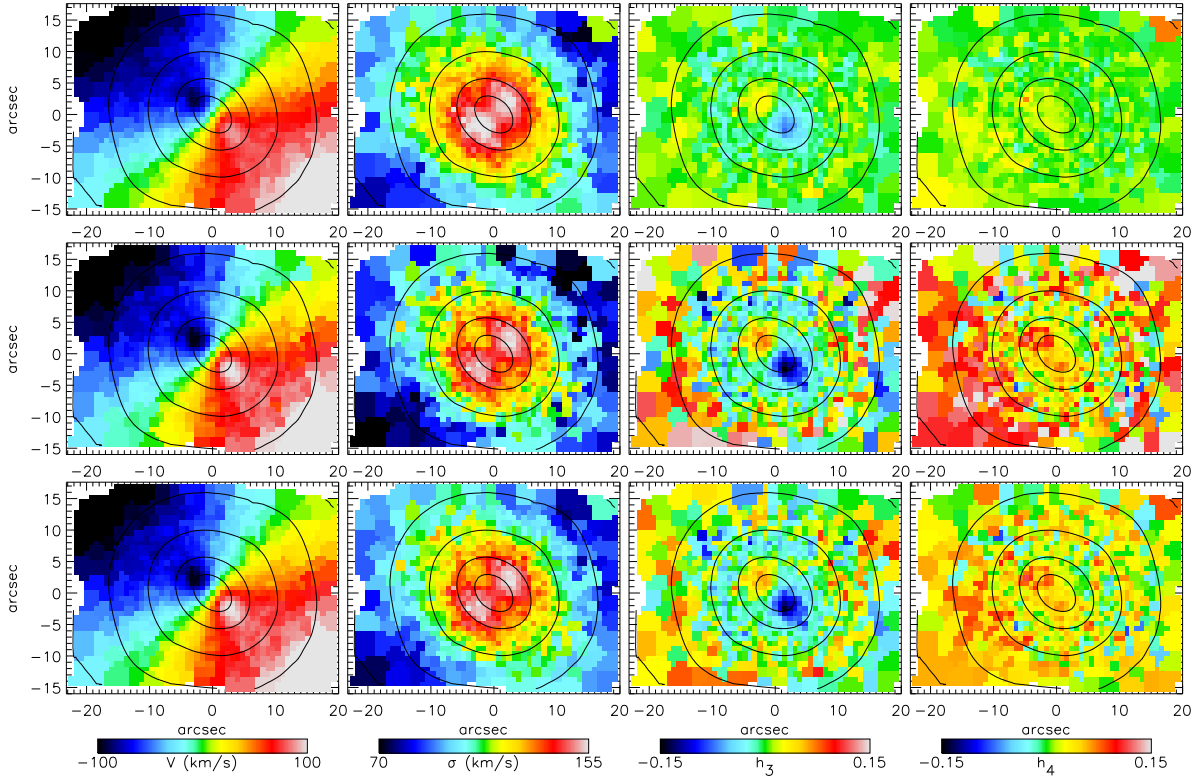}
\caption{Comparison between the Voronoi 2D-binned \citep{cap03} \sauron\ stellar kinematics of the barred lenticular galaxy NGC~3384 extracted with the different methods explored in this paper. {\em Top panels:} kinematics extracted by first fitting $(V,\sigma)$ (nonlinearly) and then expanding the $(h_3,h_4)$ parameters at the optimal $(V,\sigma)$ location. The Gauss-Hermite parameters are everywhere significantly suppressed by this method. {\em Middle panels:} kinematics obtained by fitting $(V,\sigma,h_3,h_4)$ simultaneously, without penalty. Due to the $\sigma$--$h_4$ correlation (Fig.~\ref{chi2plot}), caused by the undersampling of the LOSVD, the $\sigma$ is noisier and depressed, while $h_4$ tends to be strongly positive. The $h_3$ values fluctuates in the outer parts between large positive and negative values. {\em Bottom panels:} the kinematics measured with the new penalized PXF method (with $\lambda=0.7$). This overcomes the problems of the two previous approaches. Only the statistically significant non-Gaussian features are preserved, otherwise the solution is smoothly reduced to a Gaussian. Contours of the reconstructed surface brightness are superimposed (in 1 mag arcsec$^{-2}$ steps).\label{sauron}}
\end{figure*}

We mention here that alternative forms to Eq.~(\ref{pert}) are possible, with the requirement that the objective function has the form (\ref{penchi2}). One possibility is to include a multiplicative perturbation of the residuals:
\begin{equation}
r'_n\; =\; r_n\; ( 1 + \lambda \mathcal{D}^2).
\label{pert1}
\end{equation}
In this case the objective function becomes
\begin{equation}
\chi_{\rm p}^2\; =\; \chi^2\; (1 + 2\lambda\mathcal{D}^2 + \lambda^2 \mathcal{D}^4),
\end{equation}
and if one neglects the small term containing $\mathcal{D}^4$, this becomes again of the form (\ref{penchi2}), with $\alpha=2\lambda\chi^2$. We verified that in practice this alternative form of perturbing the residuals produces virtually the same results as using Eq.~(\ref{pert}), if the same $\alpha$ parameter is adopted. We prefer the perturbation (\ref{pert}) given its robustness against outliers, due to the use of the biweight in the computation of $\sigma(\mathbf{r})$. The form (\ref{pert1}) may be useful in implementations where the average residuals are not zero for a given $\mathcal{L}(v)$.

In Fig.~\ref{sauron} we compare the three different approaches to the recovery of the LOSVD by using them to extract the stellar kinematics from \sauron\ \citep{bac01} integral-field spectroscopic observations of the galaxy NGC~3384 \citep[see][for a description of the observations]{dez02}, which has a rather low velocity dispersion over the whole observed field. The data cube was spatially binned to a minimum $\SN=60$ using the Voronoi 2D-binning method by \citet{cap03}. The same differences that we observed using synthetic model spectra can also be seen from real data:
\begin{itemize}
\item   The Gauss-Hermite parameters are everywhere significantly suppressed when fitting $V$ and $\sigma$ first (top panels);

\item the simultaneous fit of all parameters (middle panels) tends to produce a velocity dispersion map which is noisier and has smaller values, while $h_4$ tends to be strongly positive. The $h_3$ values fluctuate in the outer parts between large positive and negative values;

\item the use of a penalty function (bottom panels) overcomes the problems of the two previous approaches. Only the statistically significant non-Gaussian features are preserved, otherwise the solution is smoothly reduced to a Gaussian.
\end{itemize}

We also applied the three methods to the extraction of the kinematics of all the 48 early-type galaxies of the \sauron\ survey \citep{dez02}. As expected from the simulations, the three techniques gave very different results only for galaxies like NGC~3384, which have a low velocity dispersion. The kinematics extracted from \sauron\ observations of galaxies with dispersion $\ga180$ \kms\ over the whole observed field provided very similar result with the three methods (though by construction not identical).

\subsection{Errors on the fitted parameters}

The availability of fast computers makes Monte Carlo simulations the preferred way to estimate measurements errors for LOSVD extraction methods. This consists of repeating the full measurement process for a large number of different realization of the data, obtained by adding noise to the original spectra.

The method discussed in this paper, like many other available methods for extracting the LOSVD, makes use of some form of filtering or penalization of the solution, thus biasing the results to suppress the noise. It may be worth emphasizing that, when computing Monte Carlo errors, one has to correct for the bias in the measurements, introduced by the noise suppression mechanism. Failure to correct for the bias in the kinematics can provide a significant underestimation of the confidence intervals of parameters determined by fitting dynamical models to the data.

One way to obtain proper errors from penalized methods consists of correcting the error estimates by increasing the percentile intervals by an amount given by the bias \citep[e.g.,][]{bee90}. In real measurements one does not know the actual bias (otherwise the measurements would have been corrected for it), so an estimate of the maximum expected bias may be used instead. This requires one to perform Monte Carlo simulations of the kinematic extraction to determine the maximum bias at different \SN\ levels and $\sigma_{\rm in}$ values.

A simpler order of magnitude estimate of the errors on the parameters may be obtained by noting that Fig.~\ref{fit_all} provides a more realistic representation of the scatter in the measurements than Fig.~\ref{penalized}, at low $\sigma_{\rm in}$ values. In the case of the penalized pixel fitting routine described in the previous section one can then obtain the errors by setting the parameter $\lambda$ in Eq.~(\ref{pert}) to a very small value before running the Monte Carlo simulations. This simple approach will generally provide a conservative estimate of the actual errors.

\subsection{The algorithm}

To summarize the discussion of the previous sections, the suggested algorithm for recovering the LOSVD parametrized as a Gauss-Hermite series is the following:
\begin{enumerate}
\item start with an initial guess for the parameters $(V,\sigma)$, while setting the initial Gauss-Hermite parameters $h_3,\ldots,h_M=0$;

\item solve the subproblem of Eq.~(\ref{model}) for the weights $(w_1,\ldots,w_K,b_0,\ldots,b_L)$;

\item compute the residuals $r_n$ from the fit using Eq.~(\ref{resid});

\item perturb the residuals as in Eq.~(\ref{pert}) to get $r'_n$;

\item feed the perturbed residuals $r'_n$ into a nonlinear least-squares optimization routine and iterate the procedure from step (2), to fit for the parameters $(V,\sigma,h_3,\ldots,h_M)$.
\end{enumerate}

\subsection{Availability}

An IDL routine implementing the algorithm described in this work is made available from the Web address \url{http://www.strw.leidenuniv.nl/$\sim$mcappell/idl/}.

\section{Conclusions}
\label{conclusion}

In this paper we addressed the problem of extracting the LOSVD of the stars, parametrized using a Gauss-Hermite series, from observed galaxy spectra. Using a method that works directly in pixel space, we compared different techniques by applying them to the same problem, with both simulated and real data, and we showed that one has to pay special attention to the extraction, when the LOSVD is not well sampled by the data, or when the \SN\ is low.

We proposed that in these situations one should apply the maximum penalized likelihood formalism to extract as much information as possible from the spectra, while suppressing the noise in the solution. We demonstrated that this leads to a fast and simple algorithm. We also discussed how Monte Carlo errors should be properly estimated. This penalized pixel fitting method is particularly useful to extract the kinematics from integral-field spectroscopic data, as for 2D maps there is no standard way to visualize errors and one wants to be able to show only the non-Gaussian features that are statistically significant. A routine that implements the ideas of this paper is made publicly available.

\acknowledgements

We thank the \sauron\ team for making the data available, and in particular Richard McDermid for fruitful discussions and testing of the method, Tim de Zeeuw, Jesus Falc\'on-Barroso, Davor Krajnovi\'c and Glenn van de Ven for commenting on the draft. We are grateful to the referee Roeland van der Marel, for useful comments that improved the presentation of this work. MC acknowledges support from a VENI grant awarded by the Netherlands Organization of Scientific Research (NWO).

\end{document}